\documentclass[prl,twocolumn,tightenlines,floatfix]{revtex4}
\usepackage{graphicx,amsfonts,amssymb,amsmath}

\newcommand{\nc}{\newcommand}
\nc{\al}{\alpha}
\nc{\ga}{\gamma}
\nc{\de}{\delta}
\nc{\ep}{\epsilon}
\nc{\ze}{\zeta}
\nc{\et}{\eta}
\renewcommand{\th}{\theta}
\nc{\Th}{\Theta}
\nc{\ka}{\kappa}
\nc{\la}{\lambda}
\nc{\rh}{\rho}
\nc{\si}{\sigma}
\nc{\ta}{\tau}
\nc{\up}{\upsilon}
\nc{\ph}{\phi}
\nc{\ch}{\chi}
\nc{\ps}{\psi}
\nc{\om}{\omega}
\nc{\Ga}{\Gamma}
\nc{\De}{\Delta}
\nc{\La}{\Lambda}
\nc{\Si}{\Sigma}
\nc{\Up}{\Upsilon}
\nc{\Ph}{\Phi}
\nc{\Ps}{\Psi}
\nc{\Om}{\Omega}
\nc{\ptl}{\partial}
\nc{\del}{\nabla}
\nc{\be}{\begin{eqnarray}}
\nc{\ee}{\end{eqnarray}}
\nc{\ov}{\overline}
\nc{\gsl}{\not\!}

\newcommand{\bi}[1]{\bibitem{#1}}
\newcommand{\fr}[2]{\frac{#1}{#2}}

\newcommand{\cp}{\;\;\slash{\!\!\!\!\!\!\rm CP}}
\newcommand{\qq}{\langle \ov{q}q\rangle}
\newcommand{\qqg}{\langle \ov{q}\ga_5q\rangle}
\newcommand{\psl}{\slash{\!\!\!p}}

\begin{document}

\title{Neutron EDM from Electric and 
Chromoelectric Dipole Moments of Quarks\\ $\;\;$ \\}

\author{Maxim Pospelov$^1$ and Adam Ritz$^2$}

\affiliation{$^1$Theoretical Physics Institute, School of Physics and Astronomy \\
         University of Minnesota, 116 Church St., Minneapolis, MN 55455, USA}
\affiliation{$^2$Department of Applied Mathematics and Theoretical
         Physics, Centre for Mathematical Sciences,\\ University of
         Cambridge, Wilberforce Rd., Cambridge CB3 0WA, UK}

\begin{abstract}
Using QCD sum rules, we calculate the electric dipole
moment of the neutron $d_n$ induced by all CP violating operators
up to dimension five. 
 We find that the chromoelectric dipole
moments of quarks $\tilde d_i$, including that of
the strange quark, provide significant contributions comparable in 
magnitude to those induced by the quark electric dipole moments $d_i$. 
When the theta term is removed via the Peccei-Quinn symmetry, 
the strange quark contribution is also suppressed and 
$d_n =(1\pm 0.5)\left\{1.1e(\tilde d_d + 0.5\tilde d_u)+1.4(d_d-0.25d_u)
\right \} $.
\end{abstract}

\maketitle

\newpage

Experimental discovery of CP violation in flavor-conserving channels 
would provide a clear indication of new physics around the electroweak scale.
This stimulates continuing experimental efforts to detect electric 
dipole moments in elementary particles and heavy atoms \cite{KL}.  
The extraordinary precision,
$d_n< 6\,\times \,10^{-26}~e$cm \cite{nEDM}, 
obtained in measurements of the electric dipole moment (EDM)
of the neutron, allows us to probe 
energy scales inaccessible in direct collider experiments. The 
information on 
fundamental CP violating parameters, such as 
CP-odd phases and masses of superpartners in supersymmetric (SUSY) models, 
is encoded in the coefficients of 
CP-odd effective operators ${\cal O}^{(i)}$, 
evolved down to 1 GeV. However, our ability to extract this 
information  depends critically on the quality of calculations 
of the neutron EDM induced by these operators.

In this Letter, we report the results of the first systematic 
analysis of $d_n$,
induced by all CP violating operators with dimension 4 and 5, within
the QCD sum rule framework \cite{sr}. The complete set $\{{\cal O}^{(i)}\}$ 
of operators at the 1 GeV scale  
includes the theta term and the electric and chromoelectric 
dipole moments of quarks 
(EDMs and CEDMs):
\be
 \de {\cal L} & = & - \sum_{q=u,d,s} \ov{q} (m_q +i\th_q\ga_5) q
               + \th_G \frac{\al_s}{8\pi}G\tilde{G} \nonumber\\
  && \!\!\!\!\!\!\!\!\!\! -\frac{i}{2}\sum_{q=u,d,s}d_q \ov{q}F\si \ga_5 q
               -\frac{i}{2}\sum_{q=u,d,s}\tilde{d}_q \ov{q}g_sG\si \ga_5 q,
   \label{deL}
\ee
where $F_{\mu\nu}$ and $G_{\mu\nu}$ are the electromagnetic and
gluonic field strength tensors. With these sources, the neutron EDM can be 
written as a linear combination
of the coefficients in Eq. (\ref{deL}),
\be
 d_{n} = d_{n}(\bar\th)+d_{n}^{\rm EDM}(d_u,d_d,d_s) 
    + d_{n}^{\rm CEDM}(\tilde{d}_u,\tilde{d}_d,\tilde{d}_s). \label{dn}
\ee
The first term $d_n(\bar\th)$ arises from the dimension four
$\th$--term in (\ref{deL}), where $\bar\th=\sum_q\th_q+\th_G$ is the physical
combination. In the absence of e.g. the 
Peccei-Quinn (PQ) compensation mechanism 
\cite{PQ}, this is the most important source term, and the calculation 
of $d_n(\bar\theta)$ within QCD sum rules was performed in 
\cite{pr2,henley}. Here we shall
present the calculation of the dimension five contributions,
$d_n^{\rm EDM}(d_q)$ and $d_{n}^{\rm CEDM}(\tilde{d}_q)$, which 
are required for phenomenological analyses. In fact, the results of 
$d_n^{\rm EDM}(d_q)$ can be extracted from calculations of the 
tensor charges of light quarks in the proton \cite{tch}. At the same time,
earlier calculations of $d_{n}^{\rm CEDM}(\tilde{d}_q)$ in
QCD sum rules \cite{CEDMsr} 
predict an unexpected suppression of CEDM contributions, 
contrasting with chiral loop estimates which 
indicate sizable effects \cite{KK}.
Therefore, the main focus of this work 
is the crucial task of calculating the contributions of CEDMs, 
which also depend on the presence or absence of 
the PQ mechanism \cite{BUP}.

The starting point for the calculation is 
the correlator of currents $\et_n(x)$ with quantum
numbers of the neutron in a background with nonzero
CP-odd sources and an electromagnetic field $F_{\mu\nu}$,
\be
 \Pi(Q^2) & = & i\int d^4x e^{ip\cdot x}
    \langle 0|T\{\et_n(x)\ov{\et}_n(0)\}|0\rangle_{\cp,F},
     \label{pi}
\ee
where $Q^2=-p^2$, with $p$ the current momentum.

In the presence of CP violating sources 
it is necessary to take
into account mixing between the neutron current and its CP conjugates.
Thus we parametrize the interpolating current $\et_n$ in the form,
\be
 \et_n & = & (j_1 +\beta j_2) + i\ep_{\cp}(i_1 + \beta i_2), \label{current}
\ee
where the two conventional neutron interpolators, 
$j_1 = 2\ep_{abc}(d_a^TC\ga_5u_b)d_c$ and
 $j_2 =  2\ep_{abc}(d_a^TCu_b)\ga_5d_c$, 
are combined with their  CP conjugates, $
 i_1 =  2\ep_{abc}(d_a^TCu_b)d_c $ and 
 $i_2  =  2\ep_{abc}(d_a^T\ga_5 Cu_b)\ga_5d_c$.
The parameters $\ep$ and $\beta$ in (\ref{current}) play rather
different roles. Firstly, $\ep$ reflects mixing induced by the CP
violating source and will be calculated below.  
Thus to linear order
in the sources,
\be
 \langle \et_n \ov\et_n \rangle & = & \langle j\ov{j} \rangle_{F,\cp}
    +i\ep_{\cp}\langle j\ov{i} + i\ov{j}\rangle_F + O(\ep_{\cp})^2, \label{mix}
\ee
where,
\be
 \ep_{\cp} & = & \frac{i}{2}\frac{\langle i_1\bar j_1 - j_1\bar
      i_1\rangle_{\cp}}{\langle j_1 \bar j_1 - i_1 \bar i_1\rangle} 
\ee

The second parameter $\beta$
reflects the existence of the two interpolators $j_1$ and $j_2$, and
would disappear from the result of an exact calculation. It can
therefore be used to optimize the convergence of the operator product
expansion (OPE) of (\ref{pi}). 
Within the sum rules formalism, one has the imperative
of suppressing the contribution of excited states and higher dimensional
operators in the OPE, and thus 
its convenient to choose $\beta$ to this end.  
We shall therefore keep $\beta$ arbitrary, and 
optimize once we have knowledge of the structure of the sum rule.

In analyzing the correlator (\ref{pi}) an additional
consideration is that when CP-symmetry is broken by a generic
quark-gluon CP-violating source, 
the coupling between the physical neutron state, described by a spinor $v$, 
and the current $\eta_n$ acquires an additional phase factor,
\be
\langle 0|\et_n|N\rangle &=& \lambda\, e^{i\al\ga_5/2}\, v, \label{coupling}
\ee
where the coupling naturally decomposes as $\la=\la_1+\beta\la_2$
which in part motivates our parametrization above, 
with the same parameter $\beta$ for the CP conjugate currents $i_1$ and $i_2$.
With regard to electromagnetic form factors, the
unphysical phase $\al$  in (\ref{coupling})
can mix electric ($d$) and magnetic ($\mu$) 
dipole moment structures and complicate the extraction of $d$ 
from the sum rule. As discussed
in \cite{pr2}, there is a unique tensor structure contributing to the neutron
double pole term which is not contaminated by this phase, namely 
$\{F\si\ga_5,\psl\}$. Therefore, it is this 
structure that we shall use to construct the EDM sum rule from
(\ref{pi}). An additional advantage of using this structure comes 
from the absence of unknown vacuum tensor polarizabilities \cite{tch} in the 
calculation of $d_n^{\rm EDM}(d_q)$. 

We now proceed to study the OPE associated with (\ref{pi}). 
To next-to-leading order in the OPE the relevant classes of 
diagrams we need to consider are shown in Fig.~1
((a), (b) and (c)). Diagrams of the form (d), although suffering
no loop factor suppression, are nonetheless suppressed due to 
combinatorial factors
and the small numerical size of $(\langle \ov{q}q\rangle)^2$ relative
to the natural scales in the problem. (This partially explains why the 
results of previous analyses \cite{CEDMsr} indicated a suppression
of $d_{n}^{\rm CEDM}(\tilde{d}_q)$).

The vacuum structure is conveniently encoded in a generalized propagator
expanded in the background field and the associated condensates.
The condensates are then parametrized in terms of various
susceptibilities defined as \cite{is}:
\be
 \!\!\!\!\!&&\langle \ov{q}\si_{\mu\nu}q\rangle_F =  \ch_q F_{\mu\nu}
           \langle  \ov{q}q\rangle;\;\;
 g\langle \ov{q}(G_{\mu\nu}^at^a)q\rangle_F 
   =  \ka_q F_{\mu\nu}\langle \ov{q}q\rangle \nonumber\\
 \!\!\!\!\!&& g\langle \ov{q} G\si q \rangle =  -m_0^2;\;\;
 2g\langle \ov{q}\ga_5(\tilde{G}_{\mu\nu}^at^a)q\rangle_F 
   =  i\xi_q F_{\mu\nu}\langle \ov{q}q\rangle,\;\;\;\;\;
\ee
and henceforth we follow \cite{is} and assume that
$\ch_q=\ch e_q$ etc., with flavor independent susceptibilities 
$\ch,\ka,\xi$. The dependence on CP violating parameters is either explicit,
e.g. $\th_q$ and $d_q$, or implicit in certain vacuum condensates
in the case of $\tilde{d}_q$ and will be discussed shortly.

The relevant contributions to the OPE are exhibited in Fig.~1(a-c), but
the explicit expressions are quite unwieldy, and we shall defer
full details \cite{pr6}, and simply 
\begin{figure}
 \includegraphics[width=8cm]{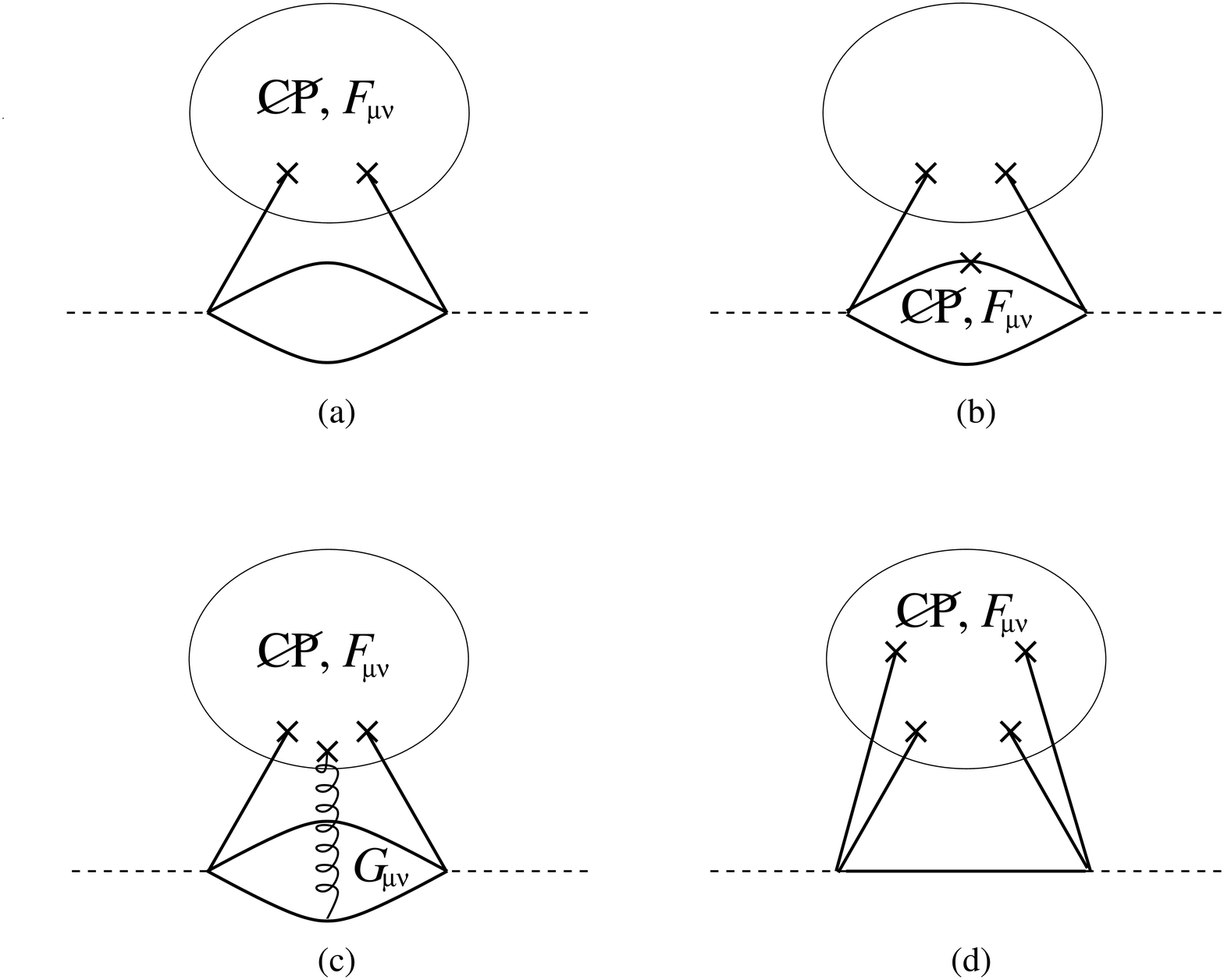}

 \caption{\footnotesize Various classes of diagrams contributing to the CP-odd structure
  $\{F\si\ga_5,\psl\}$.}
\end{figure}
\noindent
present the resulting OPE
expression. In momentum space we find
\be
 \Pi(Q^2) & = & -\frac{i\ln(-p^2)}{64\pi^2}\qq\{F\si\ga_5,\psl\}
      \left[\pi^{(\ch)}+\pi^{(\xi,\ka)}+\pi^q\right]\nonumber \\
 && \!\!\!\!\!\!\!\!\!\!\!\!\!\!\!\!\!\!+\frac{i}{16\pi^2
p^2}\ln\left(-\frac{\mu_{IR}^2}{p^2}\right)\qq\{F\si\ga_5,\psl\}
 \left[ \pi_L^v + \pi_L^d\right],
\label{ope}
\ee
corresponding to the first two nontrivial orders in the OPE. The
relevant contributions from $\bar\th$ and the CEDMs are contained
in,
\be
 \pi^{(\ch)} & = & 4(1+\beta)^2\ch_dm_dP_d-(1+\beta)^2\ch_um_uP_u
\nonumber\\
         && +2(1-\beta^2)m_*(\ch_u+\ch_d)(P_u-P_d),
\ee
and
\be
 \pi^{(\xi,\ka)} &=& \frac{1}{4}\tilde d_d \left[\al_d^+(3+2\beta+3\beta^2)
  -\al_u^+(1-\beta^2)\right. \nonumber\\
 && \!\!\!\!\!\!\!\!\!\!\!\!\!\!\!\!\!\!\!\!\!\!\!\!\!\!\!\!\!\!\!\!\!\!\!\!\!\!
     \left.-\al_d^-(1-\beta)^2\right]
   +\frac{1}{8}\tilde d_u\left[2\al_d^+(1-\beta^2)-\al_u^+
         (1+\beta)^2\right],
\ee
where we have defined $P_q=\th_q - i\qqg_{\cp}/\qq$,
$m_*=m_um_d/(m_u+m_d)$,  and
$\al^{\pm}=2\ka\pm\xi$ for concision. In obtaining this result it 
is crucial to take into account the mixing with the CP conjugate
currents in (\ref{mix}). The parameters controlling the mixing are the
unphysical phases $\th_G-\th_q$ and the quark mass difference
$m_d-m_u$. 
Here, as expected, 
the addition of the mixing terms ensures the absence of the unphysical phases
$\th_G-\th_q$. 

The quark EDM contributions are
\be
 \pi^q & = & d_d\left[10+6\beta^2\right]-d_u\left[3+2\beta-\beta^2\right].
\ee
At subleading order, terms with a logarithmic momentum dependence arise
and involve an infrared cutoff $\mu_{IR}$. The relevant contributions
are,
\be
 \pi_L^v & = & (1-\beta)^2m_de_dP_d\nonumber\\
 && \;\;\;+(1-\beta^2)m_*(e_u+e_d)(P_d-P_u)\\
 \pi_L^d & = & \frac{1}{12}(1-\beta)^2m_0^2(d_d - e_d \tilde d_d)
\ee

The next problem to address is the calculation of the vacuum 
matrix elements $P_u$ and $P_d$ in Eq. (\ref{ope}). 
These terms require the evaluation of correlators
of the form $\int d^4y\langle \ov{q}\ga_5 q(x), i\de{\cal L}(y)\rangle$, where
$\de{\cal L}$, given in Eq.~(\ref{deL}), involves 
in particular the $\th$-term and 
the color EDM sources which may be extracted from the vacuum at leading
order in the background electromagnetic field. The case of $\theta$ was 
discussed at length in \cite{pr2,Kcor,PR}. Here we shall 
concentrate on the CEDMs and  evaluate these correlators 
in chiral perturbation theory, saturating them by $\pi_0$ and $\eta_8$ 
exchange, assuming the decoupling of a heavy singlet state, and
incorporating the effect of the chiral anomaly \cite{pr4}. 
We then obtain,
\be
 m_{u(d)}P_{u(d)} \!\!&=&\!\! m_*\bar\th + 
 \frac{m_*m_0^2}{2}\left(\frac{\tilde{d}_{u(d)}-\tilde{d}_{d(u)}}{m_{d(u)}}
 -\frac{\tilde{d}_s}{m_s}\right) \;\;\;\;\;\;\;\;
\ee
where we have neglected terms of 
$O(\tilde d_{(u,d)}/m_s)$, assuming an approximate proportionality 
of the CEDMs to the quark masses, i.e.
$\tilde d_d /\tilde d_s \sim m_d/m_s\ll 1$. 

With these results in hand, we now turn to optimization of the OPE
expression (\ref{ope}). Recall that there are generally two motivated
approaches for fixing the mixing parameter $\beta$: (1) at a local 
extremum; or (2) to minimize 
the effects of the continuum and higher dimensional operators. Since
we are restricted here to only the first two orders in the OPE, we
make use of the second option, as in \cite{pr2}, and set $\beta=1$ to
cancel the subleading infrared logarithmic terms, which are ambiguous
due to the cutoff. This procedure mimics the original motivation for
$\beta=-1$ in the CP-even case \cite{ioffe}. 

On the phenomenological side of the sum rule we have
\be
 \Pi^{\rm phen} \!\!&=& \!\! \frac{i}{2}\{F\si\ga_5,\gsl p\}\left(
\fr{\lambda^2d_nm_n}{(p^2-m_n^2)^2} +
\fr{A}{p^2-m_n^2}\cdots\right)\!. \;\;\;\;\;\;\;
          \label{phenfull}
\ee
We retain here the double and single pole contributions,
the latter corresponding to transitions between the neutron and excited
states, but the exponentially suppressed continuum contribution will be 
ignored. In (\ref{phenfull}) 
$\la=\la_1+\beta\la_2$ and $A$ is an effective constant
parametrizing the single pole contributions.
A more detailed analysis will be presented elsewhere \cite{pr6}, and
we have verified that the inclusion of a continuum does not 
affect the analysis below.

After a Borel transform of (\ref{ope}) and (\ref{phenfull}), and
using $\beta=1$ as discussed earlier, 
we obtain the sum rule,
\be
 \la^2 m_nd_n+AM^2 & = &
-\frac{M^4}{32\pi^2}e^{m_n^2/M^2}\qq\nonumber\\
 && \!\!\!\!\!\!\!\!\!\!\!\!\!\!\!\!\!\!\!\!\!\!\!\!\!\!\!\!\times
        \left[\pi^{(\ch)}_{\beta=1}+\pi^{(\xi,\ka)}_{\beta=1}+\pi^q_{\beta=1}
 \right] +O(M^2),
   \label{sumrule}
\ee  
where the contributions now take the elegant form,
\be
 \pi^{(\ch)}_{\beta=1} & = & 4\left[4\ch_dm_dP_d-\ch_um_uP_u\right] \\
 \pi^{(\xi,\ka)}_{\beta=1} & = & \frac{1}{2}\left[4\tilde d_d\al_d^+ - 
  \tilde d_u \al_u^+\right] \\
 \pi^q_{\beta=1} & = & 4\left[4d_d-d_u\right].
\ee
It is remarkable that the contribution of
the $u$ quark is $-1/4$ that of the $d$ quark, which is precisely
the combination suggested by the SU(6) quark model! 

In order to analyze the sum rule (\ref{sumrule}), we 
first determine the coupling $\la$ using 
the sum rules for the tensor structures {\bf 1} and $\psl$ 
in the CP even sector (see e.g. \cite{leinweber} for a recent review).
Following \cite{pr2}, we construct two sum rules.
Firstly, (a): we extract a numerical value for $\la$ via a direct
analysis of the CP even sum rules. This analysis has been discussed before
and will not be reproduced here (see e.g. \cite{leinweber}). 
One uses $\beta=-1$, and obtains $(2\pi)^4\la\sim 1.05\pm 0.1$.
As an alternative, (b): we extract $\la$ explicitly as a function of $\beta$ 
from the CP-even sum rule for $\psl$, and substitute the result
into (\ref{sumrule}) choosing $\beta=1$. 

The conventional approach which we shall adopt here 
is to assume that $A$ is independent of $M$, and thus
the left hand side of (\ref{sumrule}) is linear in $M^2$ provided that
$\la$ is constant in the appropriate region for the Borel parameter. 
For case (b), the latter point may be verified explicitly. 
The function $\nu(M^2)$, given by
\be 
 \nu(M^2) &\equiv& \frac{1}{2[\pi^{(\ch)}+\pi^{(\xi,\ka)}+\pi^q]}
    \left(d_n+\frac{AM^2}{\la^2m_n}\right),\;\;\;\;
\ee
is then determined by the right hand side of (\ref{sumrule}).

\begin{figure}
 \includegraphics[width=8cm]{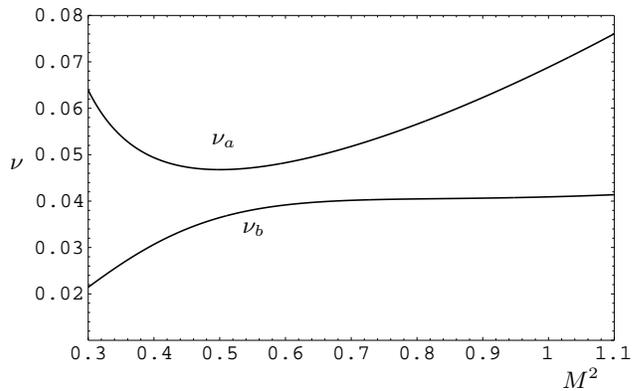}

\vspace{-3cm} \hspace{-8.5cm} $\nu$

\vspace{-0.7cm} \hspace{-3cm} $\nu_a$

\vspace{0.8cm} \hspace{-2.2cm} $\nu_b$

\vspace{1.7cm} \hspace{6.5cm} $M^2$

 \caption{\footnotesize The neutron EDM function $\nu(M^2($GeV$^2))$
is plotted according to the sum rules (a) and (b).}
\end{figure}

The two sum rules described above for $\nu_a$ and
$\nu_b$ are plotted in Fig.~2. $\nu(M^2)$ is to be interpreted as
a tangent to the curves in Fig.~2.
For numerical calculation we make use of the following
parameter values: For the quark condensate, we take
$\langle \ov{q}q\rangle = - (0.225\mbox{ GeV})^3$,
while for the condensate susceptibilities, we have the values 
$m_0^2 = 0.8 \mbox{ GeV}^2$ \cite{chival}, 
$\ch = - 5.7 \pm 0.6 \mbox{ GeV}^{-2}$ \cite{chival}, 
$\xi = - 0.74 \pm 0.2$ \cite{CEDMsr}, and
$\ka = - 0.34 \pm 0.1$ \cite{chival}.

One observes that both
sum rules have consistent extrema suggesting that our
procedure for fixing the parameter $\beta$ is appropriate. Furthermore,
the differing behavior away from the extrema implies that for consistency
we must assume $A$ to be small. One then finds 
$d_n$ as given by $\nu(M^2\sim 0.55$GeV$^2)$. 
This low scale is characteristic of
CP violating effects, and convergence of the OPE is apparently not in
danger due to the combinatoric suppression factors which arise in extracting
the CP violating sources from higher dimension operators.

Extracting a numerical estimate for $\nu(M^2\sim 0.55$GeV$^2)\sim 0.043$ 
from Fig.~2, and determining an approximate error due to
higher order corrections and the sum rule analysis, we find the result\footnote{(04/2005) v2: This 
updated expression corrects an overall factor of two error propagating from \cite{pr2}.}:
\be
 d_{n} &=& (0.4\pm 0.2) \left[\ch(4e_d-e_u)
        \left(m_*\bar\th-\frac{1}{2}m_0^2\tilde
d_s\frac{m_*}{m_s}\right) \right. \nonumber \\
&& \;\;\;\;\;\;\;\;\;\left. +\frac{1}{2}\ch m_0^2(\tilde d_d
  -\tilde d_u)\frac{4e_dm_d+e_um_u} {m_u+m_d} \right. \nonumber\\
&& \;\;\;\;\;\;\;\;\;\left.
 +\frac{1}{8}(4\tilde d_d\al_d^+-\tilde d_u\al_u^+)
+ (4d_d-d_u)\right],
\label{final}
\ee
for the neutron EDM. The CEDM contributions are significant and 
comparable in magnitude in fact to the effects induced by quark EDMs.

For most phenomenological applications, it is necessary to 
invoke a PQ symmetry, which removes the dominant contribution 
$d_n(\bar\th)$. However, while setting
$d_n(\bar\th)=0$, this symmetry
induces additional CP violating terms through linear
contributions to the axion potential \cite{BUP}. 
In particular, the axion potential
has the form $V \sim -\th^2K - 2\th K'$, where
$K=i(\al_s/8\pi)^2\langle G\tilde{G},G\tilde{G}\rangle$
is the topological susceptibility, and 
$ K'  =  i(\al_s/8\pi)\langle G\tilde{G},
     \delta{\cal L}^{\rm CEDM}\rangle $
are correlators arising from the CEDM sources. This linear
shift in the axion potential then leads to an ``induced'' $\th$-term
with coefficient (see e.g. \cite{pr4})
\be
 \th_{\rm ind} & = & -\frac{K'}{K} = \frac{m_0^2}{2}\sum_{q=u,d,s}
     \frac{\tilde d_q}{m_q},
\ee
which importantly is independent of any of the specific details of
the axion mechanism.

Including these contributions, we observe the complete 
cancellation of the term proportional to the strange quark CEDM,
and (\ref{final}) takes a remarkably simple form
\be
 d^{\rm PQ}_{n} &=& (0.4\pm 0.2) 
\left[4d_d-d_u+\frac{1}{2}\ch m_0^2( 4e_d\tilde d_d
  -e_u\tilde d_u) \right. \nonumber\\
 && \;\;\;\;\;\;\;\;\;\;\;\;\;\;\;\;\;\;\;\;\;\;\;\;\;\;\;\left.
 +\frac{1}{8}(4\tilde d_d\al_d^+-\tilde d_u\al_u^+)\right].
\label{finalPQ}
\ee
Substituting numerical values for the condensates, we obtain
\be
 d^{\rm PQ}_{n} &=& (1\pm 0.5) \frac{|\qq|}{(225 {\rm MeV})^3}
  \;\;\;\;\;\;\;\;\;\;\;\;\;\nonumber \\
  && \!\!\!\!\!\!\!\!\times\left[1.1e(\tilde d_d + 0.5\tilde d_u)
    +1.4(d_d-0.25d_u)\right].\;\;\;\;\;
\label{number}
\ee
The result for $d_n^{\rm EDM}(d_q)$ agrees well with lattice
calculations of quark tensor charges of the proton \cite{lattice}, and with the
naive quark model estimate. 
Note that an overall factor of $\qq$ combines with the light quark masses  
from short-distance expressions for $d_i$ and $\tilde d_i$ to give 
a result $\sim f_\pi^2 m_\pi^2(1+O(m_{(u,d)})$ thus reducing the uncertainty 
due to poor knowledge of the quark masses and condensates.  
Further progress in calculating $d_n$ would 
require more elaborate analysis of the 
sum rules in order to reduce the error in the overall coefficient,
while the relative coefficients of different terms in the square brackets of
(\ref{number}) are likely to remain the same. 

In conclusion, we have presented the first systematic study of 
the neutron EDM induced by all CP violating sources of dimension four
and five within QCD sum rules. We observe that the 
contributions from the CEDM sources, including that of the strange quark, 
are large -- in fact as large as those associated with the quark EDM
sources. 
This result has significant
implications for the analysis of the EDM constraints imposed on 
the supersymmetric and other models of CP violation.

\bibliographystyle{prsty}

\end{document}